\pdfoutput=1
\let\oldvec\vec
\documentclass{aa}
\let\vec\oldvec

\usepackage{graphicx}
\usepackage{amssymb,amsmath}
\usepackage{txfonts}
\usepackage{natbib}
\usepackage{hyperref}
\usepackage{rotating}
\usepackage{color, colortbl}
\usepackage{sidecap}
\usepackage{threeparttable}

\renewcommand{\vec}[1]{\mathbf{#1}}

\definecolor{LightCyan}{rgb}{0.88,1,1}

\definecolor{Gray}{gray}{0.9}

\begin{document}

   \title{Imaging Spectroscopy of Type U and J Solar Radio Bursts with LOFAR }

   \author{Hamish A. S. Reid and Eduard P. Kontar}
   \institute{SUPA School of Physics and Astronomy,   University of Glasgow, G12 8QQ, United Kingdom}

   \date{}

\abstract
{Radio U-bursts and J-bursts are signatures of electron beams propagating along magnetic loops confined to the corona.  The more commonly observed type III radio bursts are signatures of electron beams propagating along magnetic loops that extend into interplanetary space.  Given the prevalence of solar magnetic flux to be closed in the corona, it is an outstanding question why type III bursts are more frequently observed than U-bursts or J-bursts.}
{We use LOFAR imaging spectroscopy between 30--80~MHz of low-frequency U-bursts and J-bursts, for the first time, to understand why electron beams travelling along coronal loops produce radio emission less often.  Radio burst observations provide information not only about the exciting electron beams but also about the structure of large coronal loops with densities too low for standard EUV or X-ray analysis.}
{We analysed LOFAR images of a sequence of two J-bursts and one U-burst.  The different radio source positions were used to model the spatial structure of the guiding magnetic flux tube and then deduce the energy range of the exciting electron beams without the assumption of a standard density model.  We also estimated the electron density along the magnetic flux rope and compared it to coronal models.}
{The radio sources infer a magnetic loop 1 solar radius in altitude, with the highest frequency sources starting around 0.6 solar radii.  Electron velocities were found between 0.13~c and 0.24~c, with the front of the electron beam travelling faster than the back of the electron beam.  The velocities correspond to energy ranges within the beam from 0.7--11~keV to 0.7--43~keV.  The density along the loop is higher than typical coronal density models and the density gradient is smaller.}
{We found that a more restrictive range of accelerated beam and background plasma parameters can result in U-bursts or J-bursts, causing type III bursts to be more frequently observed.  The large instability distances required before Langmuir waves are produced by some electron beams, and the small magnitude of the background density gradients make closed loops less facilitating for radio emission than loops that extend into interplanetary space.}


\titlerunning{The occurrence of U-bursts and J-bursts}
\authorrunning{}

   \maketitle

\section{Introduction}\label{sec:introduction}

U-bursts and J-bursts, first reported by \citet{Maxwell:1958aa}, are believed to be the signatures of electron beams travelling along closed magnetic loops, although other alternatives exist \citep[e.g.][]{Haddock:1959aa,Takakura:1966aa,Ledenev:2008aa}.  The bursts form either an inverted `U' or `J' shape in the dynamic spectra.  The `U' shape arises from an electron beam travelling up the ascending leg of a magnetic loop through a decreasing plasma density, corresponding to a negative frequency drift rate.  The beam then travels down the descending leg through an increasing plasma density, corresponding to a positive frequency drift rate.  If the beam stops emitting just after the apex of the loop then a `J' shape is made in the dynamic spectrum.  Consequently J-bursts can be considered a subgroup of U-bursts.  

U-bursts are not observed as frequently as type III bursts, the most commonly observed solar radio burst, caused by energetic electrons travelling along magnetic loops that extend into the heliosphere.    We might expect U-bursts to be observed more often, as they are signatures of energetic electrons in closed loops and most solar coronal magnetic flux is closed in the low corona.  What are the reasons why energetic electrons are not producing radio bursts along closed flux tubes?  To answer this question, we need to understand the properties of the accelerated electrons and the magnetic loops they travel along.  In this work, we explore these properties using radio imaging spectroscopy of two J-bursts and one U-burst observed below 100 MHz.

Despite occurring less often than type III bursts, U-burst observations are not uncommon.  Over 5 years, \citet{Leblanc:1985aa} observed U-bursts between 75-25~MHz occurring on 70\% of days where activity was present (249/352) with approximately half the days consisting of isolated U-bursts and the other half consisting of U-burst groups (several U-bursts) or storms (U-bursts over a few hours to days).  Within this frequency range most U-bursts were observed as J-bursts.  The tendency of J-bursts to be observed over U-bursts highlights the difficulty of electron beams to produce radio emission in an increasing plasma density gradient.  Even when U-bursts are observed they usually have an asymmetry in their emission, with regions of positive drift rates being weaker and more diffuse than regions of negative drift rates \citep[see e.g.][]{Hughes:1963aa,Fokker:1970aa,Stewart:1977aa,Poquerusse:1984aa}.  The positive background electron density gradients hamper the growth of Langmuir waves required for radio emission, as shown in numerical and theoretical studies \citep[e.g.][]{Kontar:2001ab,Li:2011ab}.

No systematic study has been done for the heights of U-burst sources.  The spatial information from U-bursts has been predominately reported in the decametre range but with sparse frequency coverage.  The bulk of the imaging observations were taken at 160, 80, 43~MHz using the Culgoora radioheliograph \citep{Labrum:1970aa,Sheridan:1973aa,Stewart:1977aa,Suzuki:1978aa}.  The spatial observations show sources arising high in the corona, far from the associated active region at altitudes of at least 1~$R_\odot$ (solar radii) at 43 MHz.  There is a large separation in the positions of sources attributed to the ascending and descending legs of the coronal loops.  Turnover frequencies of 30~MHz or less were reported, indicating coronal loops can extend significantly higher than 1~$R_\odot$ \citep[for a review on coronal loops see e.g.][]{Reale:2014aa}.    In comparison, flux tubes that continue into the heliosphere were systematically studied by \citet{Dulk:1980aa} using over 500 type III bursts at 80 and 43 MHz.  There was a notable absence of type IIIs observed above 1.6 and 2.1~$R_\odot$ from the Sun centre for 80 MHz and 43 MHz respectively, giving the average heights in the corona for bursts detected at these frequencies of 0.6 and 1.1~$R_\odot$.  Such altitudes are higher than what standard quiet Sun density models predict for these frequencies, indicating propagation along over-dense magnetic structures.

U-bursts have been spatially observed at higher frequencies.  \citet{Aurass:1997aa} analysed 23 U-bursts at frequencies between 432~MHz and 164~MHz with the Nan\c{c}ay Radioheliograph.  Like lower frequencies, sources attributed to negative and positive frequency drift rates were imaged in spatially separated sources on loops with scales around 1~$R_\odot$.  When observed, sources with negative and positive drift rates were on opposite sides of the magnetic neutral line.  Groups of type III bursts are commonly observed around U-bursts.  When type III bursts occur within a few seconds of the U-bursts, their source locations are usually closer to the U-burst source that has a negative drift rate, indicating a common region accelerating electrons along different trajectories.  For smaller loops, only one spatial observation of U-bursts has been reported in the GHz range by \citet{Aschwanden:1992aa} who observed a series of 6 U-bursts at 1.45~GHz with the VLA.  Together with a magnetic field extrapolation they inferred a loop of height 130~Mm ($0.19~R_\odot$) and length 400~Mm.  

Velocities of the exciting electron beams have been estimated from the spectral observations of U-bursts.   Without images at multiple frequencies, some form of density model is required \citep[e.g.][]{Fokker:1970aa,Dorovskyy:2010aa} to obtain positional information.  For frequencies below 160 MHz, the velocities deduced from U-bursts are similar in magnitude to the velocities of type III bursts.  Using a series of 29 U-bursts observed below 160~MHz \citet{Labrum:1970aa} used the time associated with the width of the `U' in the dynamic spectra at twice the turnover frequency and, together with a density model, estimated average velocities of 0.25~c.  If the exciter velocity of U-bursts and J-bursts remains roughly constant then the curvature in the dynamic spectrum must correspond to a substantial decrease in the density gradient within the coronal loop, but this assumption has yet to be explored.  Recently \citet{Dorovskyy:2015aa} used the time delay between the fundamental and harmonic components of a U-burst to deduce the temperature within a magnetic loop, dependent upon the velocity of the exciting electron beam.

The time profiles of U-bursts at low frequencies have been found not to be dominated by collisional damping of plasma waves, assumed at frequencies above 1~GHz \citep{Aschwanden:1992aa,Yao:1997aa,Yao:1997ab,Fernandes:2012aa}.  For frequencies between 450~MHz to 150~MHz, \citet{Poquerusse:1984aa} found the decay time of the negative and positive drift rate branches of U-bursts to be different, with the negative drift rate branch decay time being smaller and more in line with type III decay times.  Moreover, the decay time of the positive drift rate branch does not show any statistical dependence on frequency that would be expected with collisional damping.  The lack of strong collisions allows weaker electron beams to produce U-bursts at lower frequencies than for higher frequencies.  Decay times are suggested to be caused by thermal electrons Landau damping refracted Langmuir waves from background density gradients.  Such a process has been found in simulations to be significant for propagating electron beams \citep[e.g.][]{Kontar:2001aa,Kontar:2009aa,Reid:2013aa,Ratcliffe:2014aa}.

In the present work, we use LOFAR images of J-bursts below 100 MHz to constrain the dynamics of the exciting electron beams and the density structure of the magnetic loops they travel along.  The LOFAR observations have a substantially improved frequency coverage for imaging that past observations, combined with sub-second time resolution.  We start in Section \ref{sec:observations} by giving an overview of the radio burst observations.  In Section \ref{sec:flareloop} we deduce physical values related to the exciting electron beams and the closed magnetic loop that guides the electron transport.  In the context of the current observations we then discuss the generation of U-bursts and why the observations of type III bursts are more common than U or J bursts in Section \ref{sec:discussion}.

\section{Observations} \label{sec:observations}

\begin{figure} \label{fig:spectra} \center  
\includegraphics[width=0.99\columnwidth]{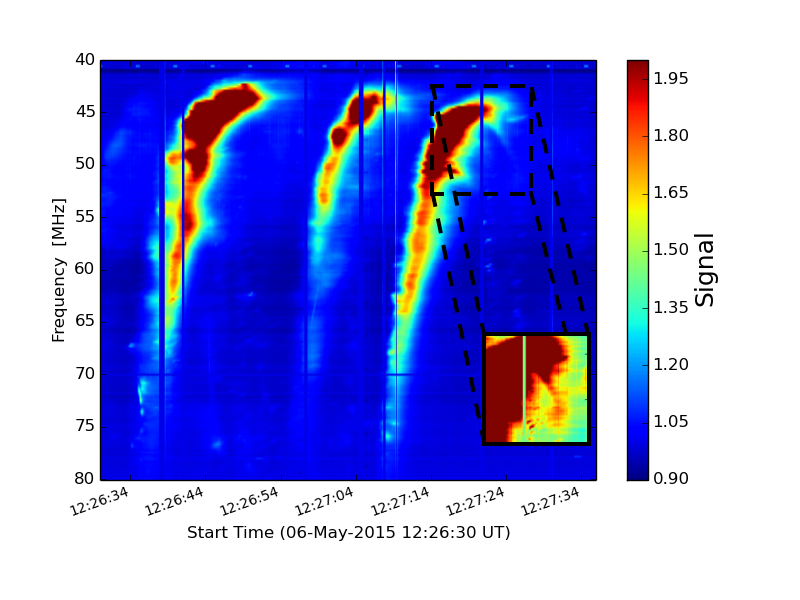}
\caption{The dynamic spectrum of two J-bursts and 1 U-burst, the three strongest bursts during a storm of bursts observed with the LOFAR LBA on the 6th May 2015.  The colour bar denotes the signal above the background frequency taken at a quiet time.  The box in the bottom right shows the U-burst when the frequency drift changes sign.  The dynamic range has been reduced to show the faint signal when the drift rate is positive.  }
\end{figure}

\subsection{Radio burst dynamic spectra}

We focus on three bursts, two J-bursts and one U-burst that were part of a storm of bursts observed around 12:30 UT on the 6th May 2015 using the Low Frequency Array \citep[LOFAR,][]{van-Haarlem:2013aa}.  The observations were made using the LOFAR low band antenna (LBA) between 80 and 30 MHz with a sub-band width of 0.192 MHz and a time integration of approximately 0.01 s.  We then integrated the time resolution to approximately 0.1~s to improve the signal to noise ratio.  The bursts started around 80 MHz and curved over in frequency space just before 40 MHz.  The spectrum of the three most intense bursts is shown in Figure \ref{fig:spectra}, normalised by a background level, where the colour axis is the log-signal above the background.  The background signal was obtained during a period of less activity between 10:50 UT and 10:55 UT.  The vertical blue lines on the spectrum are gaps in the data from interference, not normally observed in LOFAR spectra, and likely caused by lightning that was present during the observation.  The third burst is a U-burst with a weak signature that has a positive frequency drift rate between 45 and 50 MHz after 12:27:20 UT. 

\subsection{X-rays}

Just before the U and J bursts a flaring site on the North-East of the Sun produced an M2 flare at 11:50 UT, followed by a C2 flare at 12:08 UT.  At the same time as the U and J bursts, the flaring site was producing soft X-rays from heated plasma, detected by RHESSI  \citep{Lin:2002aa} below 12 keV.  A thermal fit to the soft X-ray source indicates a temperature of the plasma around 9.5 MK within the flaring site, indicated by the RHESSI contours shown later in Figure \ref{fig:movie1}.

\subsection{Burst frequency drift rates} \label{sec:driftrate}

To characterise the rise, peak and decay times for each frequency we fit the temporal profile with a function of the form
\begin{equation} \label{eqn:lightcurve}
f(t) = A_0 + A_1 \exp{\left(\frac{-t}{\tau_1}-\frac{\tau_2}{t}\right)}.
\end{equation}
Equation \ref{eqn:lightcurve} fits the rise and decay of the radio flux, where the rise and decay can occur on different time scales.  Type U, J-bursts (and type III bursts) usually have a longer decay time than rise time \citep[e.g.][]{Evans:1973aa,Poquerusse:1984aa,Krupar:2015aa}, where both time scales typically increase as a function of decreasing frequency.  The fitted function provided the peak time for each frequency, together with the rise and decay times that are taken at half the peak value (half-width half-maximum).  The fitted function also smoothed out the light curves and provided estimates during data gaps.  The duration of each burst was around 2-5 seconds for the full width half maximum (FWHM), with larger durations at the turnover frequencies around 45 MHz.  

Using the temporal fit we characterised how the bursts drifted in frequency as a function of time (drift rate, $df/dt$).  All three bursts curve in the dynamic spectrum below 47 MHz so we cannot approximate a constant drift rate across the entire burst.  Between 70-47~MHz we fit the rise, peak and decay times with a straight line and the corresponding rise, peak and decay drift rates are shown in Table \ref{tab:drift}.  The fits were found using \textbf{mpfitexy} \citep{Markwardt:2009aa} , using a temporal error of $0.05$~s ($\Delta t/2$) and a spectral error of $0.1$~MHz (half the sub-beam width).  For each burst the magnitude of the rise drift rate was greater than the peak drift rate, that was greater than the decay drift rate.  Burst 2 had a noticeably faster drift rate than burst 1 and 3 in all aspects.  The 1-sigma errors on the linear fits were at most 0.5~MHz~s$^{-1}$, with the largest error occurring from burst 2, giving a relative error of 6\%.

\vspace{20pt}
\begin{center}
\begin{table}
\centering
\begin{threeparttable}

\caption{Frequency drift rates ($\partial f/\partial t$) of the three bursts from the rise, peak and decay times.}
\begin{tabular}{ c  c  c  c   }

\hline\hline

[MHz~s$^{-1}$] 	& Rise 					& Peak 					& Decay 				\\ \hline
Burst 1 		& $-3.9\pm0.2$			&	$-3.3\pm0.1$		&	$-2.9\pm0.1$		\\
Burst 2 		& $-8.2\pm0.5$			&	$-5.7\pm0.3$		& 	$-5.2\pm0.2$		\\
Burst 3 		& $-4.8\pm0.1$			&	$-3.4\pm0.1$		& 	$-2.9\pm0.1$		\\

\hline
\end{tabular}	
    \begin{tablenotes}
      \item Notes: For frequencies 70--47~MHz.  Rise and decay times found at HWHM.  The 1-sigma errors from the linear fits are also shown.
    \end{tablenotes}
\label{tab:drift}
\end{threeparttable}
\end{table}
\end{center}

\subsection{Radio burst images}

\begin{figure} \center
 \includegraphics[width=0.99\columnwidth,trim=110 30 150 50,clip]{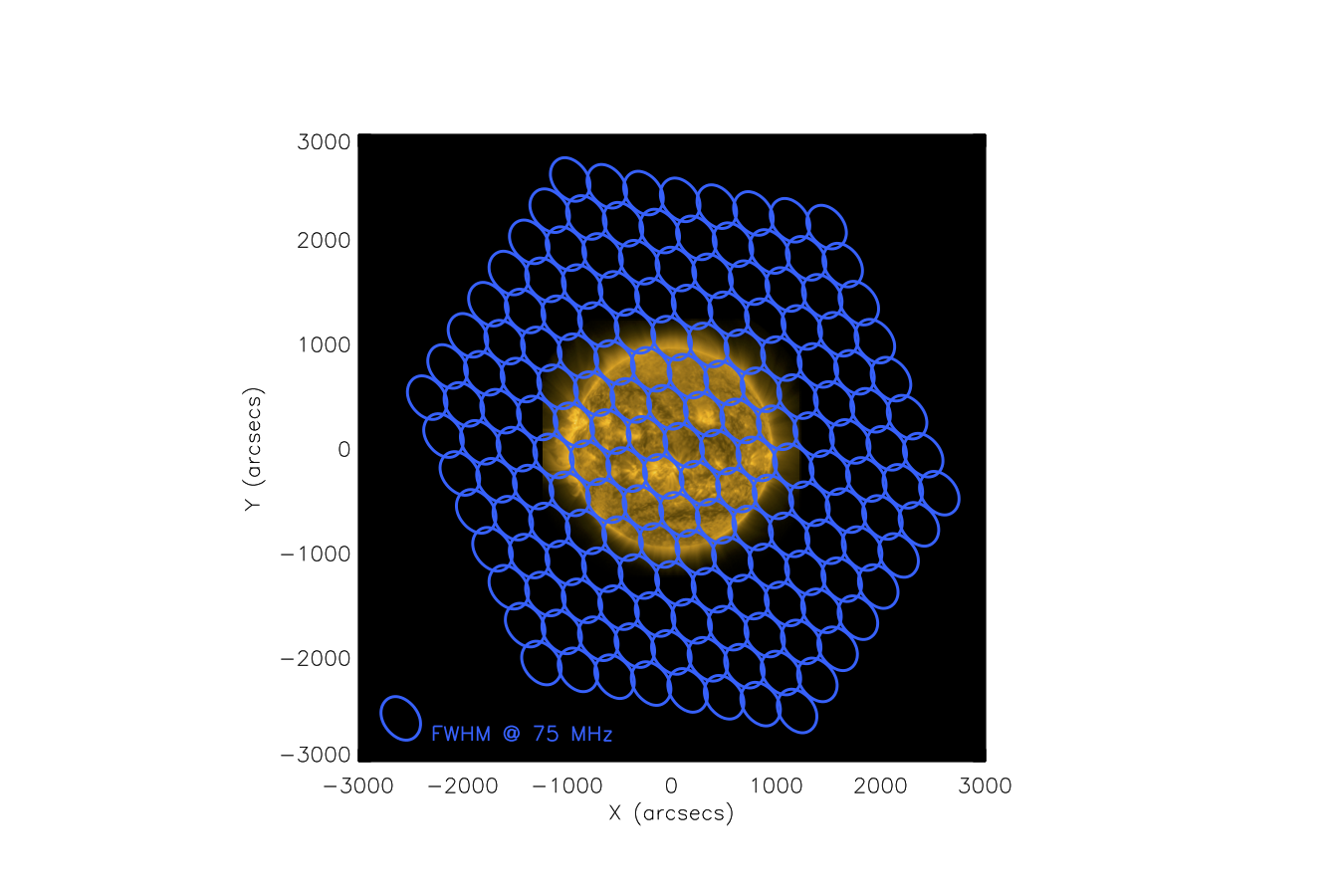}
\label{fig:mosaic}
\caption{The formation of 169 tied-array beams in a mosaic around the Sun.  Each beam displays the FWHM of the LBA beam pattern at the 6th May 2015 12:26 UT for 75 MHz.  Lower frequencies have a larger overlap.  The Sun is shown in EUV 171~\AA~from AIA.}
\end{figure}

LOFAR is able to operate in a coherent tied-array mode that involves combining the LOFAR collecting area into `array beams'; a coherent sum of multiple station beams \citep[see][for a complete description]{Stappers:2011aa}.  Each tied-array beam can be simultaneously pointed at a different part of the sky with a given right ascension ($\alpha$) and declination ($\delta$).  We used 169 tied-array beams pointed in a honeycomb pattern around the Sun.  The 169 beams allow 8 tied-array rings that cover the solar disk and the solar corona, out to a maximum of 2500~arcsecs from the disk centre.  Each tied-array beam is separated by 0.1 degree, chosen to be smaller than the FWHM of each beam at all frequencies.  The mosaic of the Sun is represented in Figure \ref{fig:mosaic}, where the FWHM of the beam shape is calculated at 75 MHz at 12:26 UT from the positions of the 24 LBA Core stations used for the observation \citep{van-Haarlem:2013aa}.

We calculate the X and Y solar coordinates from the LOFAR supplied right ascension and declination, in radians, using the offset from the solar disk centre right ascension and declination, $\alpha_s$, $\delta_s$, and a rotation about the polar angle $\theta_{pa}$, the angle of the solar north pole to celestial north, such that 
\begin{align}
 X = -(\rm{\alpha}-\rm{\alpha_s})\cos(\delta_s)\cos(\theta_{pa}) + (\delta-\delta_s)\sin(\theta_{pa}) \\
 Y = (\rm{\alpha}-\rm{\alpha_s})\cos(\delta_s)\sin(\theta_{pa}) + (\delta-\delta_s)\cos(\theta_{pa}) 
\end{align}
where $X,Y$ are then converted from radians to arc seconds.  The solar images are then generated by interpolating between the 169 mosaic points.

\begin{figure*} \center
 \includegraphics[width=0.66\columnwidth,trim=8 0 20 10,clip]{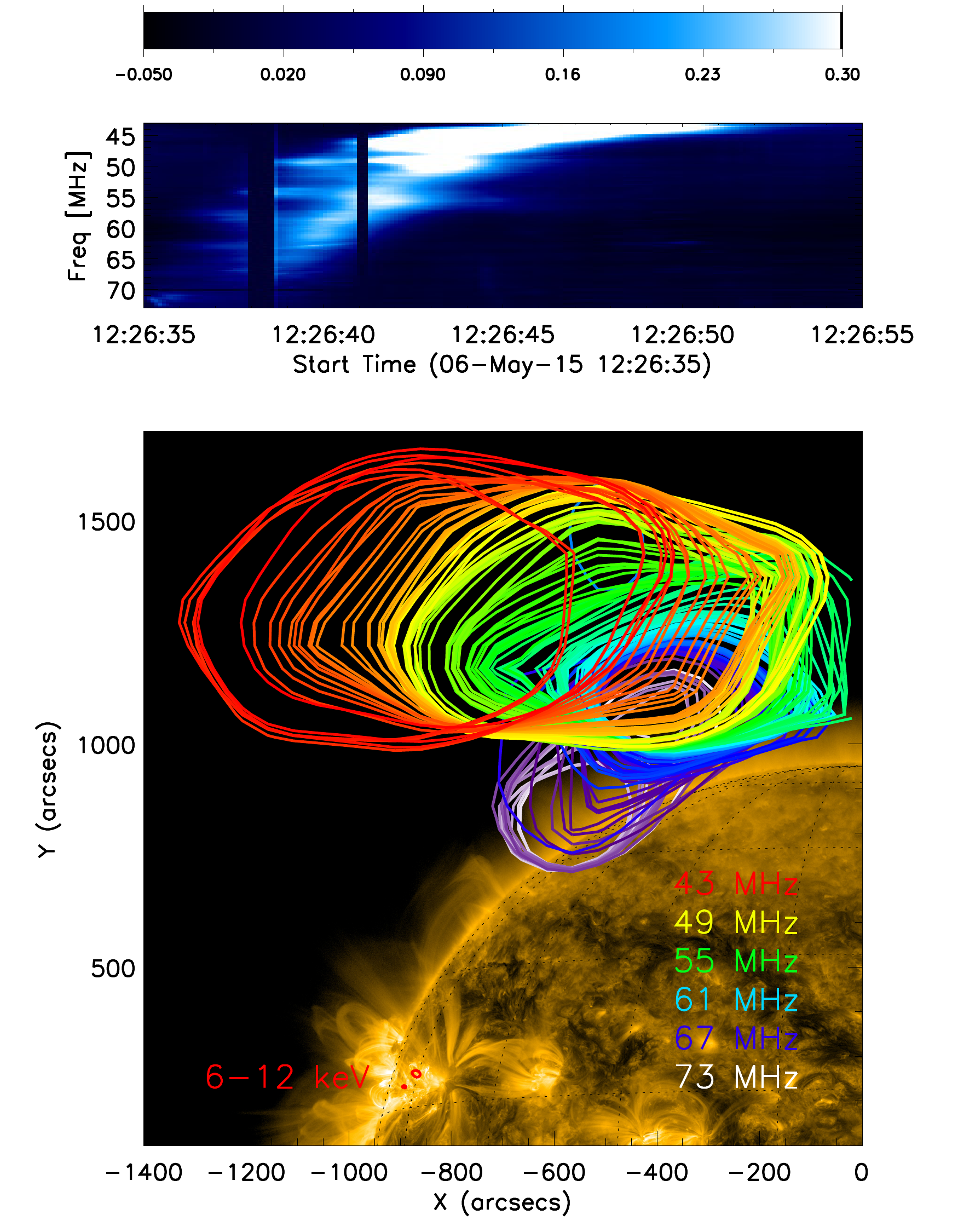}
 \includegraphics[width=0.66\columnwidth,trim=8 0 20 10,clip]{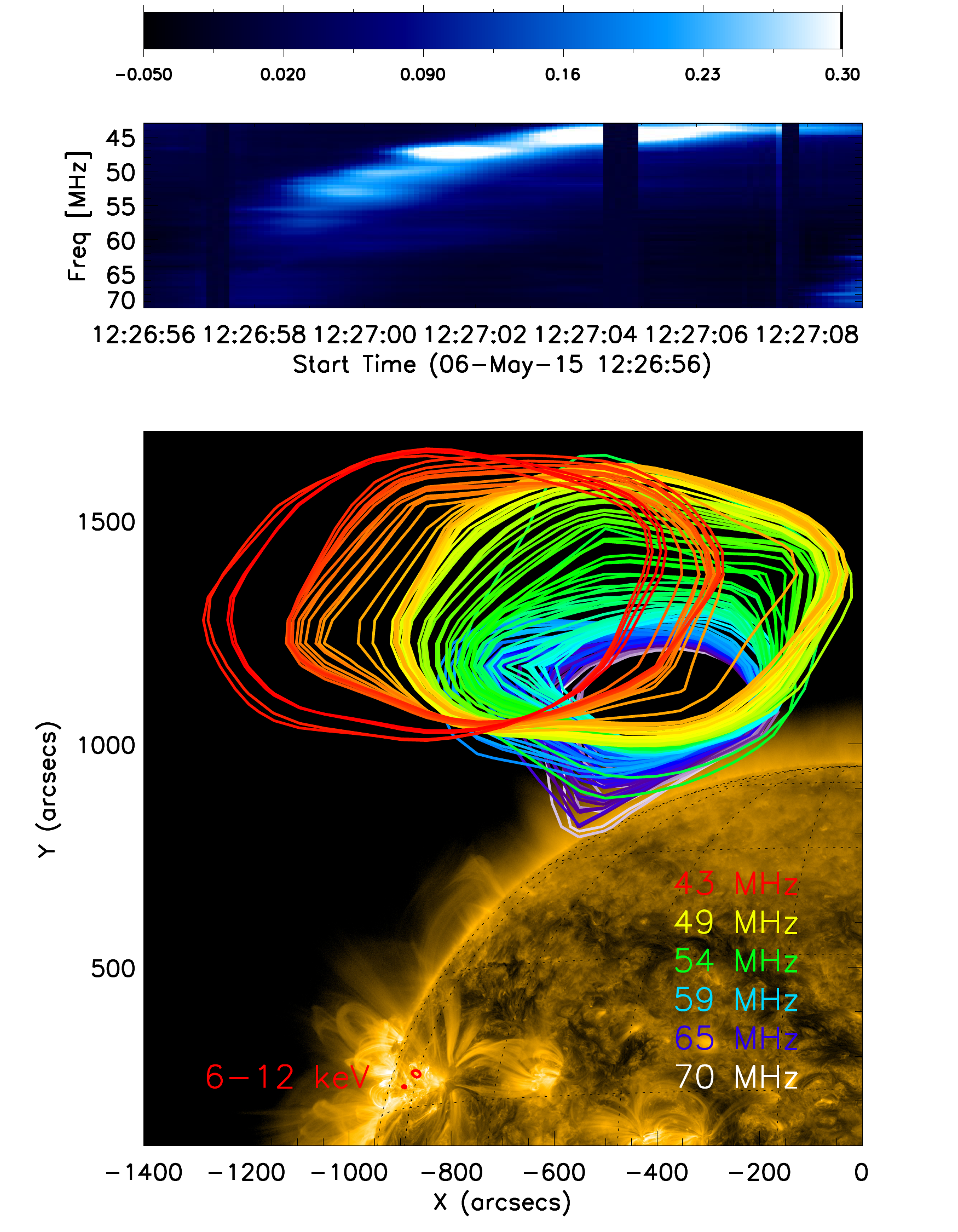}
 \includegraphics[width=0.66\columnwidth,trim=8 0 20 10,clip]{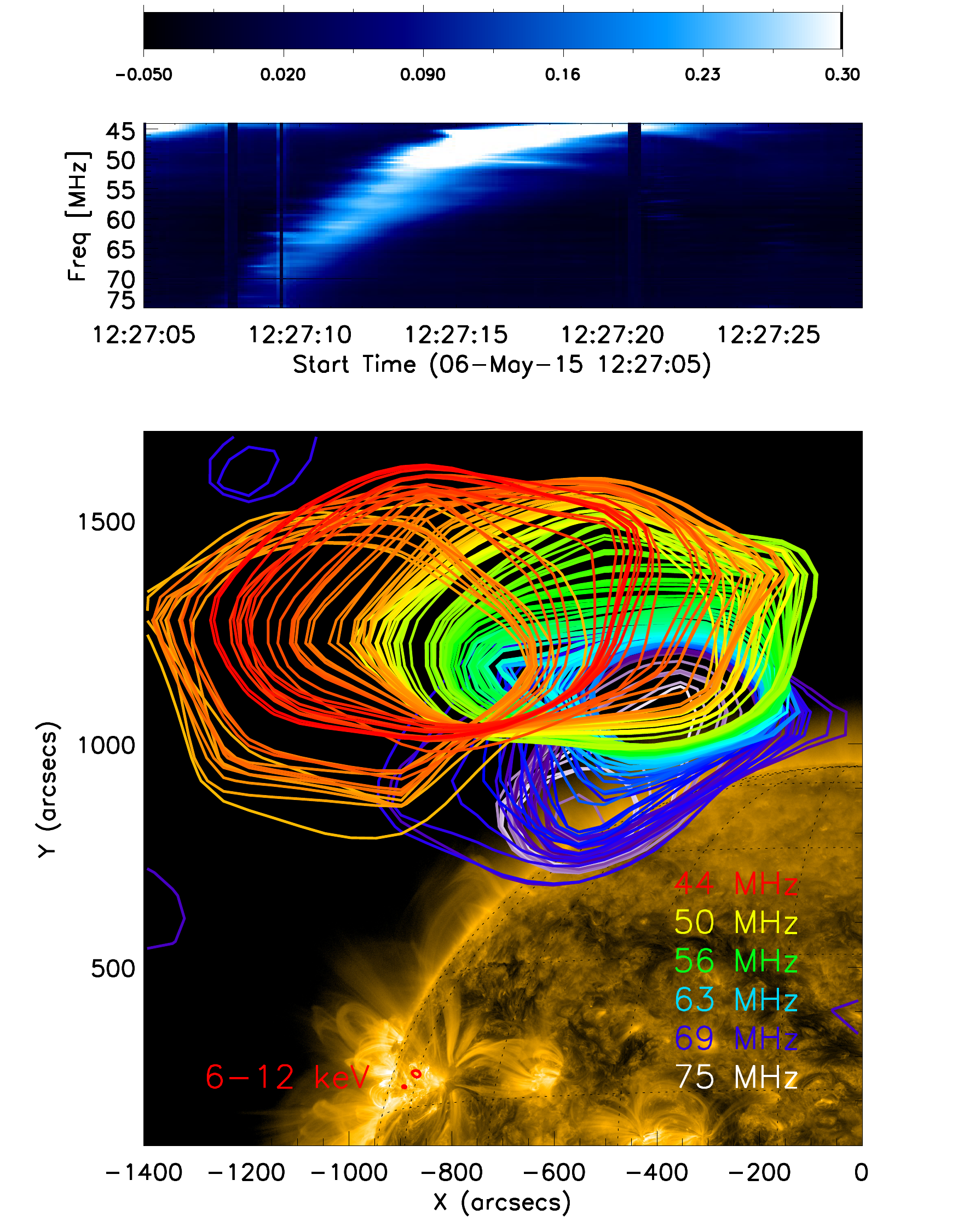}
\caption{Dynamic spectra and images of the two J-bursts and the U-burst.  Each burst is imaged at times corresponding to the peak flux in all frequencies taken from the fit using Equation \ref{eqn:lightcurve}.  The frequencies shown are the central frequency from each sub-band.  The image background is AIA EUV at 171~\AA.  The small red contours in the bottom left are the RHESSI 6-12 keV contours, imaged over the minute when the radio bursts were detected.}
\label{fig:movie1}
\end{figure*}

Images of the bursts are shown in Figure \ref{fig:movie1}.  We have plotted the $85\%$ contours for each frequency, at the time when we observe the maximum intensity in the burst.  We have used 85\% instead of 50\% as the image is still convolved with the instrument beam which will increase the size of the actual source.  The RHESSI soft X-ray source are also displayed in all three images.  The electron beam responsible for the bursts propagates in a North-West direction as it ascends the magnetic loop till around 50 MHz.  The frequency evolution then changes direction and evolves in an easterly direction.  We note that there are uncertainties in the radio source image positions from the scattering of radio waves from source to observer \citep{Kontar:2017aa}.

The third burst is a U-burst that has a faint radio source which changes sign in frequency drift rate at 12:27:20 UT.  When the positive drift rate occurs, we observe the frequency evolution to change direction and curve back towards the surface of the Sun, forming a loop shape.  The positive frequency drift rate branch is shown in Figure \ref{fig:movie1}, with the radio sources plotted at times that correspond to the maximum intensity when the frequency drift rate is positive.

\subsection{Centroid positions of radio sources}

\begin{figure}\center
 \includegraphics[width=0.99\columnwidth,trim=110 20 165 50,clip]{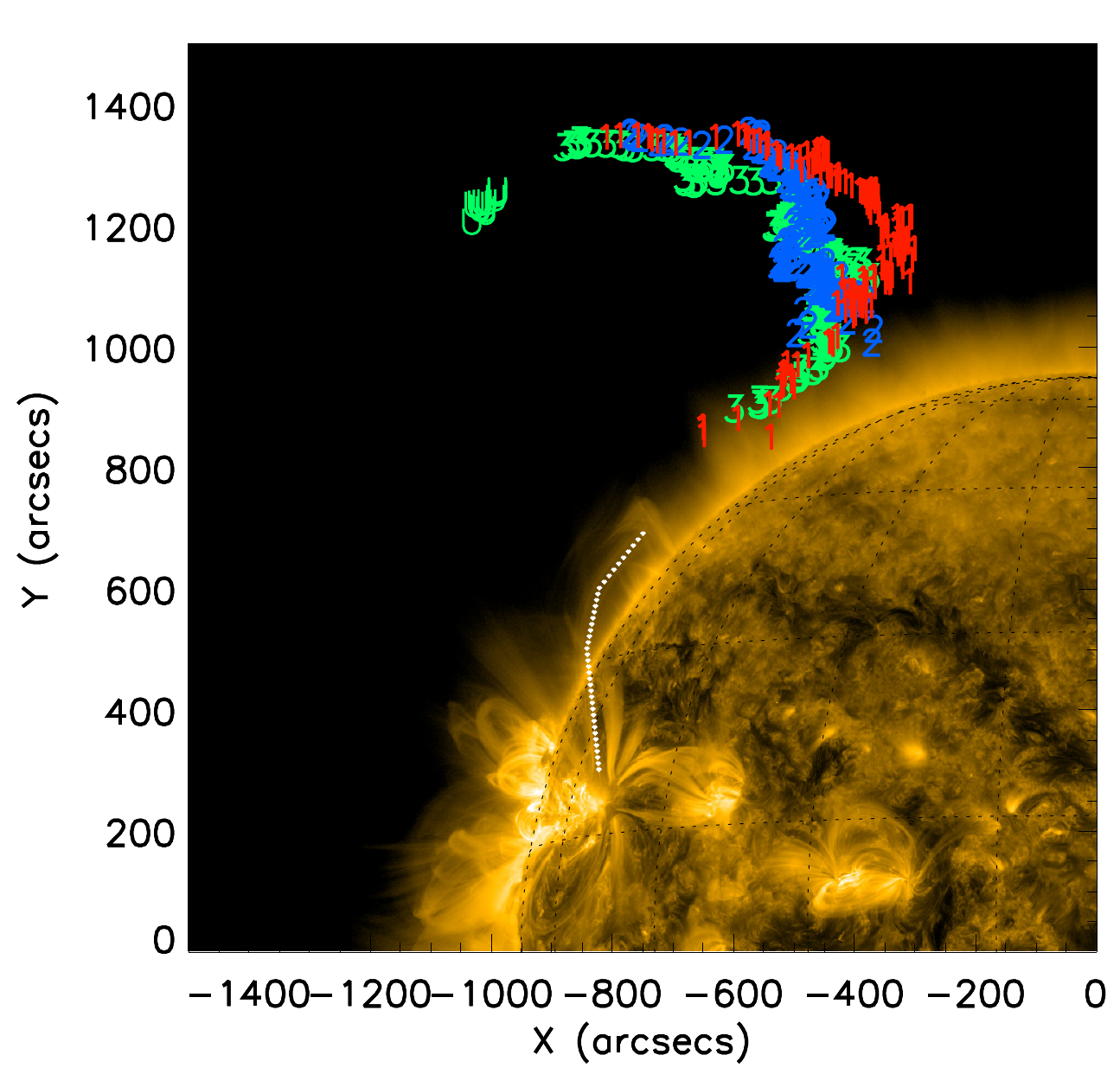}
\caption{Centroid positions of the the J-bursts (red 1,  blue 2) and the U-burst (green 3) at every 0.2 MHz.  The frequencies for each burst are the same as displayed in Figure \ref{fig:movie1}.  The positive frequency drift rate part of the U-burst is indicated as green U.  The background image is the AIA 171 \AA~.}
\label{fig:centroid}
\end{figure}

To obtain the geometry of the guiding magnetic field structure we found the centroid positions of the radio sources as a function of time and frequency.  The centroid was calculated as the first moment of the radio image, centred on the peak value of the radio image.  The size of the box to calculate the centroid was 1200 by 1200 arcsecs at 43 MHz.  We decreased the size of the box linearly in space as frequency increased, to 700 by 700 arcsecs around 70 MHz.  For every sub-band (0.192 MHz) we found the centroid at each point in time when the signal was at least 50\% of the peak signal (within the FWHM).  The sub-band centroid is then found  by averaging all the centroid positions over the time range.  The positions for each sub-band are displayed in Figure \ref{fig:centroid}.  We repeated this same process for the times when the U-burst (burst 3) had a positive frequency drift rate.  These positions are indicated in Figure \ref{fig:centroid} as a ``U''.  The positions show what looks like the electrons travelling along a magnetic field returning to towards the Sun.

The centroid positions for all three bursts agree well, showing what appears to be the geometry of a magnetic loop high in the corona.  The positions from burst 1 between 50--60~MHz are slightly farther west than burst 2, 3 by around 100 arcsecs.  This is higher than the standard deviation over the centroid positions in the x-axis for all three bursts, around 40 arcsecs.  The shift could be caused by electron acceleration occurring on slightly different field lines.

EUV observations observed by AIA in 171~\AA~(1~MK plasma) show what appears to be a loop structure curving up from the active region, close to where the X-ray source is present, towards the radio sources from the bursts.  We have indicated the path of this loop structure in Figure \ref{fig:centroid} with a white dashed line.  One end of the flare loop extends northwards around $(-800,300)$~arcsecs and curves Westward, becoming invisible around $(-725,800)$~arcsecs.  The loop structure is stable in time, being observable in 171 \AA~during all three bursts.

\section{Physical parameters of the electron beam and the coronal loop } \label{sec:flareloop}

\subsection{Coronal loop fitting using radio source positions}

Limited spectral resolution in previous low-frequency radio images has hampered radio burst observations being used to analyse large coronal loop geometry.  Using the LOFAR images at each frequency we approximate the coronal loop that guides the propagation of the radio exciter by fitting the centroid positions.  From the centroid positions and the positions of the AIA loop we have used a 4th order polynomial in $x$ to approximate the loop path from the active region to the top of the loop.  Lower order polynomials were not able to create a loop-like structure.  We approximate the top part of the loop using a 4th order polynomial fit in $y$ to the radio centroids above $y=1100$~arcsecs, including the centroids denoted with ``U'' for J-burst 3.

Using the latitude of the soft X-ray source as the base of the loop, we found the distance along the loop $l$.  The loop apex is around 1.5~$R_\odot$ along the loop whilst the altitude of the loop apex is 1~$R_\odot$.  For each LOFAR frequency between 70 to 45 MHz we found the closest point of $l$ to the centroid positions to construct the radial dependence of frequency, $f(l)$.

The positions of the highest frequencies correspond to where the electron beam started to produce radio emission.  Using observations from the French radiospectrometer ORFEES above 130 MHz, we do not observe any radio signature at higher frequencies.  The distances from the RHESSI soft X-ray source give starting heights of $0.6~R_\odot$ whilst the distance along our fitted magnetic loop give $0.8~R_\odot$.

We considered how projection effects changed $l$.  If we assume that the loop is semicircular such that the height is $\pi/2$ times the length, then a rotation about the y-axis of around 30 degrees satisfies this criterion.  Such a small rotation only increases the values of $l$ at the LOFAR frequencies by around 0.1 $R_\odot$.

\subsection{Exciter Velocities}

Exciter velocities of radio bursts are typically calculated assuming either fundamental or second harmonic emission to convert $f\rightarrow n_e$.  A density model is assumed to convert $n_e\rightarrow l$, where $l$ is the path of the electron beam.  This allows the drift rate $df/dt$ of the radio burst to be related to $dl/dt=v$.  With images from LOFAR we can directly observe the motion of the radio sources with near-continuous coverage.  We can thus estimate the exciter velocities without an assumed emission mechanism or any density model.

From the rise, peak and decay times of the bursts as a function of frequency, found in Section \ref{sec:driftrate}, we approximate the motion of the front, peak and back of the electron beam between 70--45~MHz.  There is a significant increase in the separation of the radio source centroids ($df/dl$ increases) for the lowest frequencies.  Correspondingly, the drift rate $df/dt$ of the J-bursts decreases dramatically in magnitude for the same frequencies.  The combination of these two effects means that the motion of the electron beam $dl/dt=v$ can be fit with a constant velocity between 70--45~MHz, and the velocities are given in Table \ref{tab:velocities}.  The linear fit was obtained using \textbf{mpfitexy} \citep{Markwardt:2009aa}, using a temporal error of $0.05$~s ($\Delta t/2$) and a spatial error of $180$~arcsecs (half the beam separation).  The velocities associated with the rise times are larger than the velocities associated with the peak times that are in turn larger than the velocities associated with the decay times.  The errors on the velocities are at most $0.03$~c for burst 2, giving a relative error of 15\%.  Additional uncertainties on the derived velocities will arise from our model of $f(l)$ and from positional errors relating to the scattering of the radio waves in the corona between source and observer.  The motion of the sources in frequency and time did not suggest a non-constant velocity.  Deceleration of the exciter has been inferred from the frequency and time of type III observations below 1~MHz \citep[e.g.][]{Fainberg:1972aa}, suggesting a deceleration of $12.3\pm0.8~\rm{km~s}^{-2}$ \citep{Krupar:2015aa}.

Electron beams producing Langmuir waves are not mono-energetic but occur over a range of velocities.  From beam-plasma theory \citep{Melnik:1999aa} the velocity of the beam-plasma structure at one point in space moves at the average velocity of the participating electrons $(v_{max}+v_{min})/2$.  Landau damping from a background Maxwellian plasma means that the minimum velocity must be above at least $4~v_t$ and so for a 1 MK plasma we assume $1.56\times10^9~\rm{cm~s}^{-1}$.  The velocities from Table \ref{tab:velocities} give an energy range of beam electrons between 0.7--43~keV for the highest velocity and 0.7--11~keV for the lowest velocity.  The low energy range agrees with the absence of significant hard X-ray observations above 12~keV by RHESSI.  Whilst we must treat the derived energies with care, they give an indication that the electrons responsible for the observed radio emission have low-energies around 10~keV and the spread in relevant energies does not change dramatically between the front and back of the electron beam.

\begin{center}
\begin{table} \centering
\begin{threeparttable}

\caption{Velocities derived from the radio source centroids and the rise, peak and end times.}
\begin{tabular}{ c  c  c  c  }

\hline\hline

 [c]	&	Rise			& 	Peak 			&	Decay 		\\ \hline
Burst 1 &	$0.22\pm0.03$ 	&	$0.18\pm0.02$ 	&	$0.16\pm0.02$	\\ 
Burst 2 &	$0.23\pm0.03$ 	&	$0.20\pm0.03$ 	&	$0.16\pm0.02$	\\
Burst 3 &	$0.18\pm0.02$ 	&	$0.16\pm0.02$ 	&	$0.13\pm0.02$	\\

 \hline
\end{tabular}
    \begin{tablenotes}
      \item Notes: For frequencies 70--45~MHz.  The 1-sigma fitting errors are also shown.
    \end{tablenotes}
\label{tab:velocities}
\end{threeparttable}
\end{table}
\end{center}

\subsection{Inferred background density model} \label{sec:densmodel}

\begin{figure} \center
 \includegraphics[width=0.99\columnwidth]{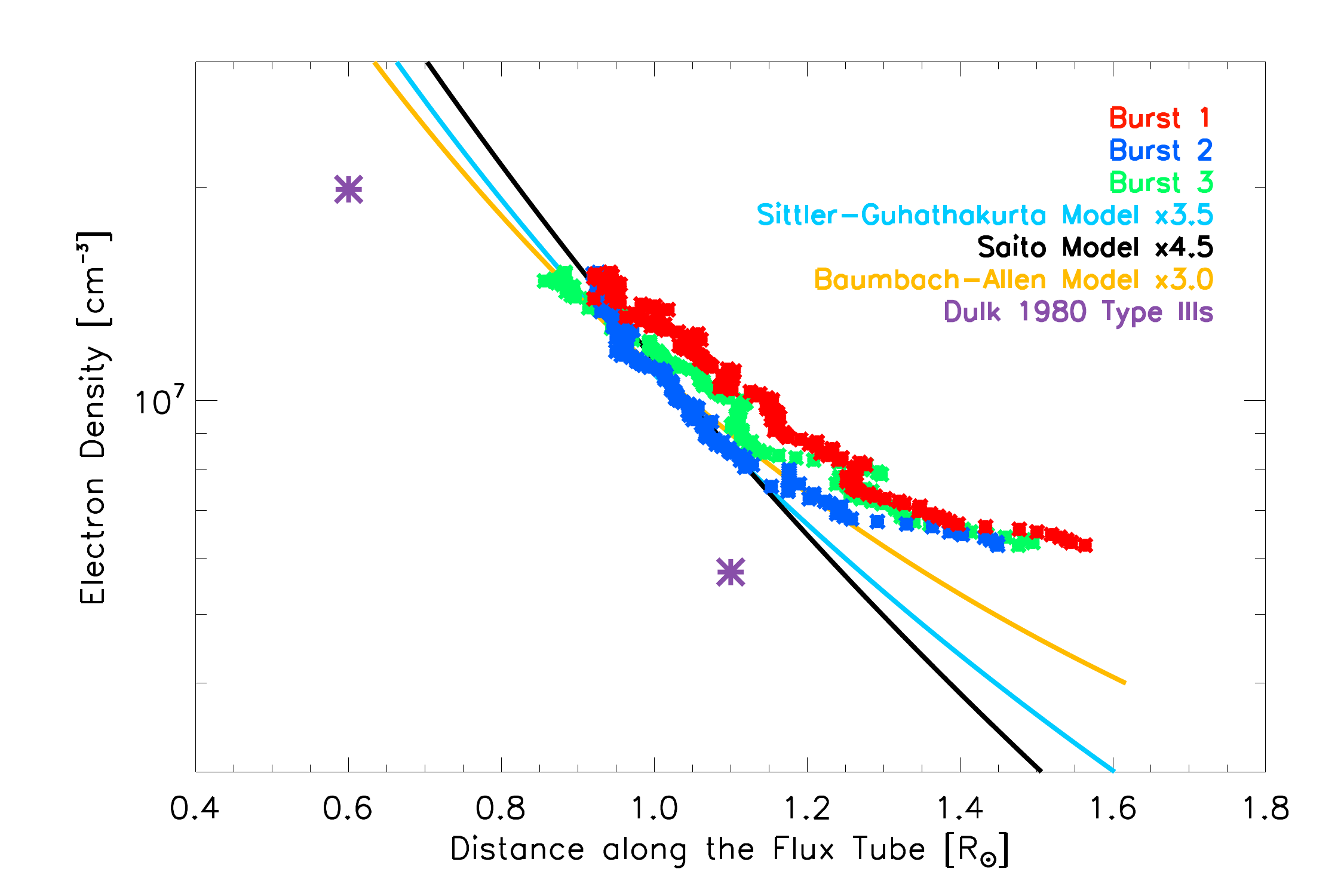}

\caption{Distances along the flux tubes for the background electron density, derived from fitting the three radio bursts in this study.  Also shown are three empirical density models multiplied by constant factors to intersect the densities derived from the bursts.  The two radial distances deduced from 600 type III bursts by \citet{Dulk:1980aa} at the corresponding densities for 80~MHz and 43~MHz are shown as purple stars.}
\label{fig:dens_fit}
\end{figure}

We used our model of $f(l)$ to obtain a density model for the coronal loop.  From the distance of the emission to the Sun and the diffuse characteristic of the radio emission, we infer second harmonic plasma emission \citep{McLean:1985aa}.  As such the radio emission originates in plasma with a background electron density that has a plasma frequency of half the radio frequency.  Explicitly we can infer the loop electron density $n_e(l)$ using
\begin{equation}\label{eqn:density}
n_e(l) = \frac{m_e(\pi f(l))^2}{4\pi e^2}
\end{equation} 
where $m_e$ is the electron mass, $e$ is the electron charge and $f(l)$ is the radio frequency in Hz along the magnetic loop.  To model $n_e(l)$ we use the loop fit of the negative frequency drift rate part of each burst.  Figure \ref{fig:dens_fit} shows the resultant plot of $n_e(l)$.

To compare our density models to other well-known empirical models, we have plotted the Sittler-Guhathakurta model \citep{Sittler:1999aa}, the Saito model \citep{Saito:1977aa} and the Baumbach-Allen model \citep{Allen:1947aa}, all multiplied by a different constant so they roughly overly the densities corresponding to the chosen LOFAR frequencies.  All three empirical density models had to be multiplied by around half an order of magnitude; the empirical models represent open flux tubes in the quiet Sun whilst the J-bursts and U-bursts arise from a closed flux tube in the active Sun.  The magnitude of the density gradient is smaller for the loops than the empirical models, particularly at the highest altitudes.  This is related to the empirical density models occurring at different altitudes from the surface of the Sun because of the curvature in the loop.

The positions of $n_e(l)$ are slightly higher than those derived from \citet{Dulk:1980aa} using type III bursts at 80 MHz and 43 MHz.  We might expect the distance along the magnetic structures derived from type III bursts to be smaller than those derived from U-bursts and J-bursts, as type III bursts relate to open flux tubes whilst U-bursts and J-bursts will always be associated with closed flux tubes.  However, there is an inherent uncertainty in the origin that we used for our derived loop and in the AIA data we used to fit the base of the loop.  A direct line from the highest frequency radio source to the RHESSI source is around 0.6~$R_\odot$.  This would shift the points in Figure \ref{fig:dens_fit} by 0.2~$R_{\odot}$ to the left making them more in-line with the points derived from \citet{Dulk:1980aa}, particularly at 80 MHz.

Similar to \citet{Paesold:2001aa,Saint-Hilaire:2013aa} we can fit $n_e(l)$ using an exponential density model of the form
\begin{equation}\label{eqn:exp_dens}
n_e(l) = A_n\exp(-r(l)/r_n) 
\end{equation}
where $l$ is the distance along the loop in cm.  The characteristic scale $r_n$ varies between $3$--$4\times10^{10}$~cm but the fit does not accurately capture the decrease in the density gradient $dn(l)/dl$ at the lower densities (see Figure \ref{fig:dens_fit}).  If instead of the distance along the loop $l$, we use $r$ the radial distance of $l$ from the solar surface (altitude) assuming a semicircular loop, with the lowest frequency occurring at the loop apex, then we obtain a characteristic scale of $r_n$ that varies between $1$--$2\times10^{10}$~cm.  This fit better captures the decrease in $dn(l)/dl$ at lower densities.  The characteristic scale of the exponential fits are much higher than previously deduced by \citet{Paesold:2001aa,Saint-Hilaire:2013aa} of $r_n=7.5\times10^9$~cm and $r_n=3.2\times10^9$~cm respectively for density models associated with flux tubes that guide type III bursts.  The large value of $r_n$ when we assume the electron density exponentially decreases in height highlights the reduced magnitude of the density gradient in the magnetic loop compared to density models obtained using studies of type IIIs.  We note that \citet{Aschwanden:1992aa} found a upper limit of the scale height of $3.7\times10^{10}$~cm for the small magnetic loop inferred from a U-burst at GHz frequencies.

\section{The occurrence rate of U-bursts} \label{sec:discussion}

Analysing two J-bursts and one U-burst using LOFAR allowed us to image the loop geometry that guided the propagation of the electron beams.  There are uncertainties in the radio source positions, particularly from the scattering of the radio waves as they propagate from the source to the observer.  What is clear is the loop shape traced by the changing frequencies of the U-bursts and J-bursts reinforces the theory of electron beams in closed loops exciting U-bursts and J-bursts \citep{Maxwell:1958aa,Fokker:1970aa}.

We now address the question of why less U-bursts are observed than type III bursts.  The prevalence of magnetic loops confined to the low solar corona would naively lead one to expect more U-bursts, a signature of accelerated electrons propagating along these loops, rather than type III bursts, a signature of accelerated electrons propagating along magnetic field lines that extends into the interplanetary medium.  We use the deduced variables from the bursts in Section \ref{sec:flareloop}, namely the average velocity of the exciter, the density gradient of the loop and the starting height of the three bursts using the distance from the co-temporal soft X-ray source.

\subsection{Langmuir wave instability distance} \label{sec:inst_dist}

Before an electron beam can start to produce coherent radio bursts, the distribution function must become unstable to Langmuir wave production (i.e. a positive gradient in velocity space) \citep{Ginzburg:1958aa}.  From X-ray observations \citep[e.g.][as a review]{Holman:2011aa} and in situ observations at 1~AU \citep[e.g.][]{Krucker:2007aa} electron beams are injected with an initial power-law velocity distribution with a negative spectral index.  This is initially stable to Langmuir wave production as the velocity gradient is negative.  Through propagation, a beam can become unstable via time-of-flight velocity dispersion.  For U-bursts to be emitted, a magnetic loop must be large enough for propagation effects to make an electron beam become unstable before it reaches the apex of the loop.  Radio emission from the region exhibiting a positive frequency drift rate is usually weaker and more diffuse \citep[e.g.][]{Poquerusse:1984aa,Aurass:1997aa}, and if the beam emits only in the descending leg of the magnetic loop then it will be observed as a reverse type III burst.  For the bursts observed in this study, their starting height is at least 0.6~$R_\odot$ away from the active region and 0.8~$R_\odot$ along the magnetic loop that we fitted to the radio centroids.  

The distance an electron beam travels before an instability occurs (instability distance) depends largely upon the properties of the accelerator.  From comparisons between type III bursts and hard X-rays \citep{Reid:2011aa,Reid:2014aa} we found that the injected spectral index of the electron beam and the longitudinal extent of the acceleration region are important in governing the instability distance.  Assuming an acceleration region size of $d=10$~Mm, height $h_{acc}=50$~Mm and a high electron beam spectral index of $\gamma=16$ in velocity space (the bursts are not very intense), for an instantaneous injection we expect the radio emission to start at $h=d\gamma+h_{acc}=0.3~R_{\odot}$.  The discrepancy between this height and the 0.6~$R_\odot$ that we observe can be explained by a higher altitude acceleration region, or low beam densities that require a higher beam-background electron density ratio before the relevant wave-particle interactions are significant \citep{Reid:2011aa}.

The temporal injection profile of electrons also governs the instability distance.  The temporal injection plays a significant role if $v\tau_{inj}\geq d$ where $v=5\times 10^{9}~\rm{cm~s}^{-1}$ is the average velocity of the electrons derived from the J-bursts and $\tau_{inj}$ is the injection time.  Assuming $d=10$~Mm, the injection time can be as low as 0.2 seconds before it significantly extends the instability distance, creating a starting height of $h=(d+v\tau_{inj})\gamma+h_{acc}$.  An injection time $\tau_{inj}=0.5$~s provides a starting height of $h=0.64$~$R_\odot$, similar to what we observed for the J-bursts.

For smaller magnetic loops (e.g. 0.1~$R_\odot$ from base to apex), to keep $h-h_{acc}$ small enough for an electron beam to generate Langmuir waves in the ascending leg, $\tau_{inj}$ must be around 0.1 seconds or less, assuming the velocities of 0.15~c found in this study and a velocity spectral index of 10.  This is similar to the time-of-flight delay measured from hard X-rays \citep[see][as a review]{Holman:2011aa}.  The acceleration region would have to be small, and situated near the base of the loop.  Whilst not unrealistic parameters for electron acceleration, if $\tau_{inj}$ is much longer or the beam propagates at a faster velocity, then $h-h_{acc}$ becomes larger than half the loop length.  Slower electron beams nearer the thermal velocity have a smaller instability distance but are more susceptible to scattering, especially in the high densities of small magnetic loops.

\subsection{Langmuir wave instability timescale}

\begin{figure*} \center
 \includegraphics[width=0.99\textwidth]{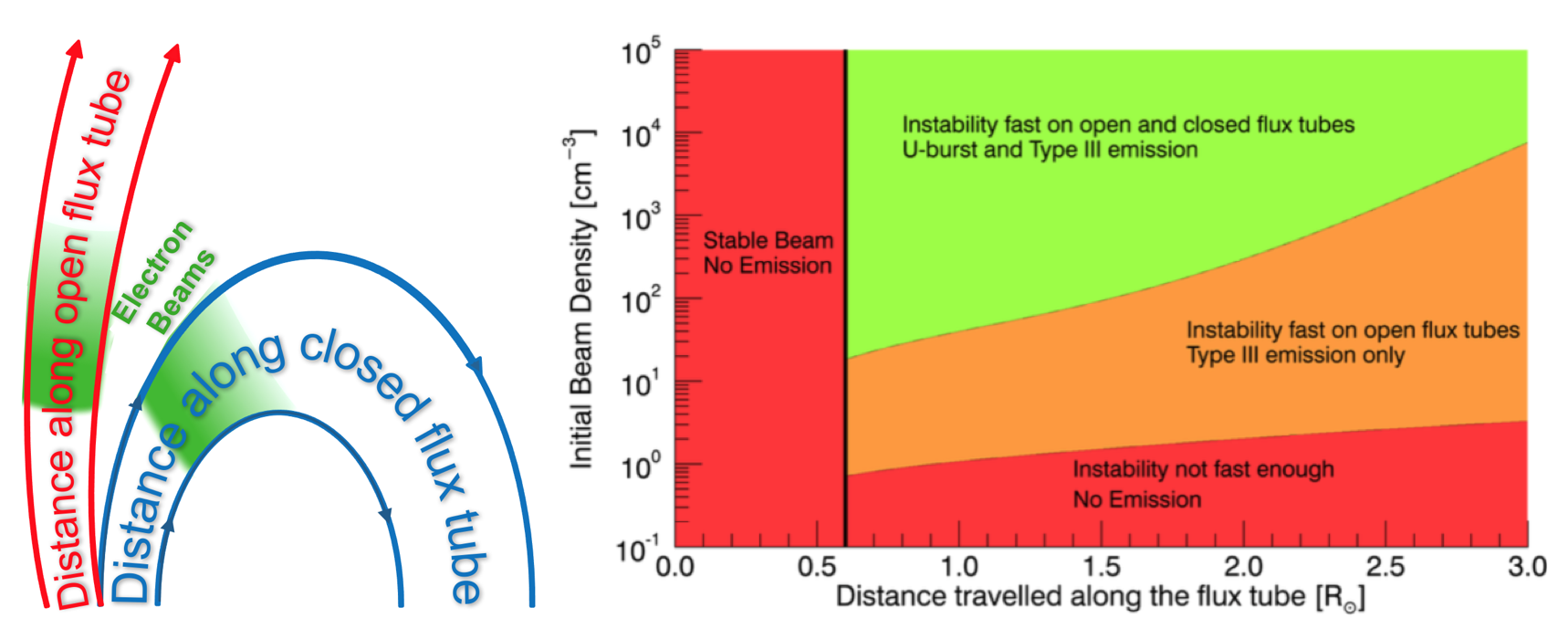}
 \caption{Left:  cartoon showing two flux tubes, one open (red) and one closed (blue) to the heliosphere with propagating electron beams (green).  Right: regions of electron beam instability to Langmuir waves for closed flux tubes (U-bursts) and open flux tubes (type III bursts).}
\label{fig:tql}
\end{figure*}

The generation of Langmuir waves requires an electron beam to be unstable but it also requires the characteristic time of wave growth (quasilinear time, $\tau_{ql}$) to be small.  The quasilinear time \citep{Vedenov:1963aa,Kontar:2001ab} can be presented as a function of distance
\begin{equation}
\tau_{ql}(l)=\frac{n_e(l)}{\omega_{pe}(l)n_b(l)}\propto \frac{n_e^{0.5}(l)}{n_b(l)},
\end{equation}
where $n_e(l)$, $n_b(l)$ are the background and beam densities respectively and $l$ is the path the electron beam travels along.  If the quasilinear time is too large then, despite a positive gradient in velocity space, significant Langmuir waves growth will not occur.

From the quasilinear time, Langmuir waves will grow faster when $n_e(l)$, the background density, is low.  The J-burst and U-burst observations inferred a low-magnitude loop density gradient that will keep $n_e(l)$ higher at distances farther from the acceleration site, unfavourable for Langmuir wave growth.  In comparison, the higher magnitude density gradient inferred from type III bursts observations means that $n_e(l)$ is lower at distances farther from the acceleration site, favouring Langmuir wave growth.  

Langmuir waves will grow faster when $n_b(l)$, the beam density, is high.  The magnetic flux tubes that guide the propagation of the electron beam have an increasing cross-section as the magnetic field strength decreases with distance from the solar surface.  The increasing cross-section of the flux tube causes the beam density to decrease as a function of distance.  As the beam density decreases, the quasilinear time $\tau_{ql}(l)$ increases, reducing the amount of Langmuir waves produced.

What quasilinear times are too large?  Electrons in the beam must be present at one point in space for longer than the quasilinear time if waves are to be generated.  For any point in space $l$, the electron beam will be present for a propagation time $\tau_p$ that depends upon the length $d$ and velocity $v$ of the beam such that $\tau_p=d/v$.  Langmuir waves can grow at one point in space if $\tau_{ql}(l)/\tau_p < 1$.

Velocity dispersion of the beam, caused by the spread of velocities within the electron beam, will elongate the beam as the faster electrons outpace the slower electrons.  The elongating beam will increase in beam length $d$ and decrease in beam density $n_b$.  The increasing beam length will increase the propagation time whilst the decreasing beam density will increase the quasilinear time, both at a similar rate.  When Langmuir wave growth becomes significant, the quasilinear diffusion of electrons causes the resulting beam-plasma structure to travel at a constant speed, reducing both effects.  As such, when comparing the ratio $\tau_{ql}(l)/\tau_p$, we ignore the effects of the elongation of the electron beam, caused by velocity dispersion.

For electron beams that would produce U-bursts and type III bursts, we compared the different regimes of wave growth using the ratio $\tau_{ql}(l)/\tau_p(l)$ for different beam densities.  We assumed a beam velocity of $5\times 10^{9}~\rm{cm~s}^{-1}$ and an instability distance of 0.6~$R_\odot$.  We have represented the density model of an open flux tube using the Baumbach Allen model, corresponding to a type III producing scenario.  The density model for a closed flux tube is represented using the density scale height of $1.48\times10^{10}~\rm{cm}$, deduced from the J-burst observations. The expansion of the magnetic field for a closed flux tube was assumed to be similar to the field expansion of a magnetic dipole such that $B(r)\propto r^{-3}$.  We model the corresponding decrease in the electron beam density as a function of distance from the acceleration region $l$ using $n_b(l)=n_b [l_0/(l+l_0)]^{3}$ with $l$ in solar radii and $l_0=3.5\times10^9~\rm{cm}=0.05~R_\odot$ used in \citet{Reid:2013aa}.  The expansion of a magnetic field for an open flux tube, in the absence of solar gravitation and the outward acceleration by high coronal temperature, can be modelled as radial \citep{Parker:1958aa}, accurate above some distance $r=b$, such that $B(r)\propto r^{-2}$.  Whilst for the frequencies measured by LOFAR, $r$ might be less than $b$, we assume radial expansion as a limiting case.  We therefore model the decrease in the electron beam density for the open flux tube using $n_b(l)=n_b [l_0/(l+l_0)]^{2}$.

Figure \ref{fig:tql} demonstrates the different regimes of wave growth and subsequent radio emission.  Before 0.6~$R_\odot$ of beam propagation, the electron beam is stable, regardless of the beam density and background density model (see Section \ref{sec:inst_dist}), so no radio emission will be observed.  The instability distance of 0.6~$R_\odot$ is dependent upon electron beam parameters [e.g. spectral index, size, time][]\citep{Reid:2011aa,Reid:2013aa,Reid:2014aa} and so will vary from beam to beam.  At distances greater than 0.6~$R_\odot$, there are three regimes:
\begin{itemize}
\item If the initial electron beam density is too low, the quasilinear time will always be larger than the propagation time ($\tau_{ql}/\tau_p > 1$).  Consequently, no Langmuir waves will be produced and no radio emission will be observed.  
\item For moderate initial electron beam densities travelling along open flux tubes, the large magnitude of the background density gradient, and the reduced field expansion makes the quasilinear time smaller than the propagation time ($\tau_{ql}/\tau_p < 1$), exciting Langmuir waves required for type III radio emission.  The same electron beams travelling along closed flux tubes have higher associated quasilinear times ($\tau_{ql}/\tau_p > 1$) because of the smaller magnitude background density gradient and the increased field expansion, and so no U-bursts are observed.
\item High initial electron beam densities result in a smaller quasilinear time than propagation time ($\tau_{ql}/\tau_p < 1$) within both open and closed flux tubes, leading to both type III and U-burst radio emission.  For U-burst emission from closed flux tubes, the required initial beam density increases at greater distances from the acceleration region, and will likely contribute to the decreased radio emission observed in the positive drift rate region of U-bursts, and the existence of J-bursts.  
\end{itemize}
If the expansion of the magnetic field is similar between open and closed flux tubes, the difference in the magnitude of the density gradient will still cause a discrepancy between which electron beams generate radio emission, albeit at a reduced effect.

\section{Summary} \label{sec:conclusion}

We have used LOFAR imaging spectroscopy to analyse two solar radio J-bursts and one U-burst from a storm of bursts that occurred shortly after a large flare at 12:00~UT on the 6th May 2015.  The J-bursts and U-bursts are a signature of accelerated electrons travelling along a closed magnetic loop.  The images from LOFAR between 80 to 40 MHz with a fine frequency coverage indicated a loop-shaped structure extending from the flaring active region where an X-ray source was present.  The bursts allow a large part of the magnetic loop to be visible at altitudes not dense enough for EUV or X-ray imaging.  The U-burst also showed faint radio emission originating from the descending leg of the magnetic loop.

A fit to the radio centroids finds a loop with an altitude of approximately 1~$R_\odot$ and a length around 1.5~$R_\odot$ from the bottom to the apex of the loop.  Starting heights for the radio emission were between 0.6--0.8~$R_\odot$.  The magnetic loop model was combined with the frequency evolution in time to estimate exciter velocities without requiring the common assumption of a coronal density model or emission mechanism.  We found velocities between 0.13~c and 0.23~c, indicating electron energy ranges from 0.7--11~keV and 0.7--43~keV respectively.  Velocities associated with the front the beam were faster than those associated with the back of the beam.  The low-energies that we found for the electron beams exciting the radio bursts agrees with a lack of X-ray response from RHESSI in the lower atmosphere above 12~keV.

We used the inferred magnetic loop from the U and J bursts to estimate a density model, assuming second harmonic emission.  The distances along the magnetic structures in the model were higher than deduced from standard quiet Sun density models and from type III bursts, although there are uncertainties in the radio positions from scattering effects of the light from the source to the observer.  The magnitude of the spatial gradient in the derived density model was smaller than standard density models of the quiet Sun, scaled to agree with the densities from the magnetic loop.  The magnitude of the density gradient decreased dramatically near the apex of the magnetic loop and the densities could be fit by an exponential density model with scale height between 1--2$\times10^{10}$~cm.

One conjecture for U-bursts is that multiple bursts in quick succession could be caused by an electron beam being mirrored by the focussing magnetic field at each end of a magnetic loop \citep[e.g.][]{Aurass:1997aa}.  The three bursts in our study show radio sources all tracing a similar path through the corona, making it likely that each burst was produced by a separate electron beam.  This result agrees well with the observation of \citet{Aurass:1997aa} that accompanying type III bursts appear close to the negative frequency drift rate branch of the U-bursts.  Type N bursts are believed to be signatures of electrons mirrored at the base of magnetic loops \citep{Caroubalos:1987aa,Wang:2001ab,Kong:2016aa}.  The properties of the third leg in an N-burst must be significantly different from the first leg, as any mirrored electron beam would likely undergo a decrease in concentration.

We used the above parameters to address the outstanding question of why U-bursts are not observed more frequently than type III bursts, considering the prevalence of closed loops in the corona.  Electron beams injected as a power-law require an instability distance before they become unstable to Langmuir waves.  The high starting heights from the J-bursts and U-burst highlighted that small loops can have insufficient distances in the ascending leg for instabilities to occur.  The low magnitude of the density gradient in the closed loops causes the quasilinear time (characteristic time of wave-particle interaction) to be higher than for open loops, reducing the growth rate of Langmuir waves required for radio emission.  We conclude that, whilst electron acceleration will frequently take place at the base of closed coronal loops, the geometrical and plasma parameters must be favourable if radio U-bursts or J-bursts are to be observed.

\begin{acknowledgements}
We thank the anonymous referee for their helpful comments and suggestions.  We acknowledge support from the STFC consolidated grant 173869-01.  Support from a Marie Curie International Research Staff Exchange Scheme Radiosun PEOPLE-2011-IRSES-295272 RadioSun project is greatly appreciated.  This work benefited from the Royal Society grant RG130642.  This paper is based (in part) on data obtained with the International LOFAR Telescope (ILT). LOFAR \citep{van-Haarlem:2013aa} is the Low Frequency Array designed and constructed by ASTRON.  We thank the staff of ASTRON, in particular Richard Fallows, for making the LOFAR observations possible.  
\end{acknowledgements}

\bibliographystyle{aa}
\bibliography{/Users/hamish/Documents/Papers/ubib}

\end{document}